\documentclass{elsart}
\usepackage[square,sort,comma,numbers]{natbib}
\usepackage[dvips]{graphicx}
\usepackage{amssymb}

\begin{document}
\begin{frontmatter}

\title{A 4$\pi$ BaF$_2$ detector for (n,$\gamma$) cross section
       measurements at a spallation neutron source}

\author[FZK]{M.~Heil}, 
\author[FZK]{R.~Reifarth}, 
\author[LANL]{M.M.~Fowler}, 
\author[LANL]{R.C.~Haight}, 
\author[FZK]{F.~K{\"a}ppeler} ,
\author[LANL]{R.S.~Rundberg}, 
\author[LANL]{E.H.~Seabury}, 
\author[LANL]{J.L.~Ullmann}, 
\author[LANL]{J.B.~Wilhelmy}, 
\author[FZK]{K.~Wisshak} 
\address[FZK]{Forschungszentrum Karlsruhe, Institut f\"ur Kernphysik, P.O. Box 3640, D-76021 Karlsruhe, Germany}
\address[LANL]{Los Alamos National Laboratory, Los Alamos, NM 87545, USA}

\maketitle

\begin{abstract}
The quest for improved neutron capture cross sections for
advanced reactor concepts, transmutation of radioactive wastes as
well as for astrophysical scenarios of neutron capture
nucleosynthesis has motivated new experimental efforts based
on modern techniques. Recent measurements in the keV region
have shown that a 4$\pi$ BaF$_2$ detector 
represents an accurate and versatile instrument for such studies.
The present work deals with the potential of such a 4$\pi$ BaF$_2$
detector in combination with spallation neutron sources, which offer
large neutron fluxes over a wide energy range. Detailed Monte Carlo
simulations with the GEANT package have been performed to
investigate the critical backgrounds at a spallation facility,
to optimize the detector design, and to discuss alternative solutions.
\end{abstract} 
\end{frontmatter}

\section{Introduction}

The development of intense spallation neutron sources
\cite{LBR90,AAA99} has made possible the measurement of neutron capture
cross sections on submilligram samples \cite{UHF99,WFH99}.
This is a necessary development for neutron cross section
measurements of rare isotopes, especially radioactive isotopes.
The new capabilities of these neutron sources are applicable to
basic research on the nature of the neutron capture process,
applied research in support of the accelerator transmutation of
radioactive waste (ATW), research in support of nuclear
astrophysics research, as well as other applications. The neutron
energy range over which cross sections need to be measured for our
specific applications is below 1 MeV, where the capture cross
sections are generally large, where the ATW schemes often have
large neutron fluxes, and where astrophysical nucleosynthesis
takes place for producing s-process isotopes and in the later 
stages of the r and p
processes. 

These applications make certain demands on the overall performance of the
neutron time-of-flight spectrometer, including the $\gamma$-ray detector array.
The spectrometer must be able to make accurate measurements of neutron capture
on small targets, e.g. submilligram, because the isotopes are inherently
difficult to produce and in some cases have a very high specific radioactivity.
High detector efficiency is desired for good counting statistics, for minimizing
corrections for missed events, and for measuring
the total energy emitted as $\gamma$-rays. With good energy resolution and a
well defined efficiency response function,
true capture events can be distinguished from the random $\gamma$-rays emitted
from radioactive targets and from many sources of background capture events
simply by demanding that the detected total energy of the gamma rays be equal 
to the Q-value of the desired capture reaction plus the kinetic energy of
the incident neutron.

Of particular importance also is the sensitivity of the detector to neutrons,
both
those scattered from the target and target backing as well as background from
other sources. With pulsed monoenergetic
sources, the background from neutrons scattered by the sample can often be
separated cleanly from the capture $\gamma$-ray events by time-of-flight (TOF)
techniques. Even if the detector captures the scattered neutrons, these
events come much later than $\gamma$-rays from neutron capture on the sample.
For example, the flight time of a 100 keV neutron over a 10 cm path is 22.9
ns, entirely adequate for distinguishing it from a capture $\gamma$-ray.
Unfortunately, this technique cannot be applied to neutron TOF spectrometers
with so-called ``white'' neutron sources because of the wide range in neutron
energy (subthermal to several hundred MeV). With a white neutron source,
scattered neutrons can arrive at the
detector at the same time as capture $\gamma $-rays from slower neutrons. A very
instructive example of this has been given by Guber et al. \cite{GSK97b}. The
design must therefore compensate
for this shortcoming in order to make measurements of the accuracy previously
achieved with monoenergetic or nearly monoenergetic sources \cite{WVK98a}.

Despite this disadvantage of white neutron sources, they have very strong
advantages.  Resonance properties can be studied thoroughly as the neutron
energy sweeps through the resonances; the energy range is continuous from
subthermal to well over 1 MeV; and techniques are well developed for
measurements of neutron flux and various sources of backgrounds.

White neutron sources are based on electron linear accelerators or
medium or high energy proton accelerators, which produce neutrons
by spallation reactions. Although a proper comparison
depends on many parameters, spallation sources typically have
intensities greater than those of electron linacs by several
orders of magnitude in the region below a few tens of keV. A
comparison of the LANSCE spallation neutron source with those
based on electron linacs has been given previously \cite{MiW90}. A
further advantage is that the ``gamma flash'' of electron linacs
is significantly reduced in spallation sources, although the
latter do produce a large flux of high energy (up to several
hundred MeV) neutrons. Although present spallation sources,
designed mostly for condensed matter research, have a worse time
spread than electron linac sources, this is not an
inherent limitation, and future sources \cite{AAA99} could be
designed to reduce this difference. Because of the advantages of
spallation neutron sources, we focus our discussion on them,
although many characteristics of the detectors in our design would
be useful also at other white neutron sources.

To effectively use the new capabilities of spallation neutron sources, new
techniques and new detectors will be required. In this paper, we discuss the
potential for a highly segmented array of BaF$_2$ detectors surrounding the
sample in which the capture cross section is to be measured. This detector
array is segmented so that it can handle the radiations from the radioactive
decay which might have the intensity of a Curie (3.7$\times$10$^{10}$ Bq) or more. We
prefer an array
that acts as a calorimeter in the sense that the total $\gamma$-ray energy
from a capture event is measured; this approach is very useful in
separating capture events on the sample from those on other materials in the
region.

In this paper we first discuss the general requirements
for a 4$\pi$ detector.  We then compare the results of a GEANT
simulation to the measured parameters of the Karlsruhe 42-detector
array to validate the calculational approach.  Finally, we apply 
GEANT to simulate a proposed 162-element
BaF$_2$ array for use at the LANSCE spallation neutron source, and
a similar array proposed for use at CERN.

Other 4$\pi$ arrays of scintillators have been developed, e.g. for
research in light and heavy ion reactions \cite{HSD79}. Originally, these detector
arrays have relied on NaI(Tl) or CsI(Tl or pure) as the scintillation materials.
Our designs are based to a large extent on the geometries of 
those arrays. In
neutron capture reaction studies, an array of 8 elements of BaF$_2$ was
developed previously for use at LANSCE \cite{Koe99}, and BaF$_2$ has been
used also in single or small arrays with lower efficiency \cite{GSK97b}. The
4$\pi$ array of 42 detectors at Karlsruhe \cite{WiK84,WGK89,WGK90a} in use
with a low energy neutron source has proved to be extremely productive in
the field of cross section measurements for nuclear astrophysics, and we
base much of our design considerations on that instrument.

With the simulations reported here, we believe that greater confidence can
be had in the choice of design of the new detector systems.

\section{Requirements of a 4$\pi$ detector}

\subsection{General aspects}

The principle of using a 4$\pi$ detector with high $\gamma$-ray
efficiency and reasonably good resolution is a complete
detection of the prompt $\gamma$-ray cascade emitted in a capture
reaction. This concept offers the advantage of obtaining a clear
signature for capture events via the sum energy of $\gamma$
cascades, which reflects the binding energy of the captured
particle. Provided that the detector is segmented into a
sufficiently large number of independent modules, valuable
additional information on event multiplicities and hit patterns
can be obtained as well. Accordingly true
capture events can be reliably distinguished from
$\gamma$-backgrounds which are inherent to neutron experiments.

In this respect 4$\pi$ arrays differ completely from the simpler
approach based on the detection of single capture $\gamma$-rays.
This category uses detectors with $\gamma$-ray efficiencies which
are linearly increasing with $\gamma$-energy. It includes
Moxon-Rae type detectors \cite{MoR63,JaK96b}, where this feature
is approximated by a $\gamma$-ray converter followed by a thin
plastic scintillator, as well as the commonly used and confusingly
named  ``total energy'' detectors.  The latter design is based on
C$_6$D$_6$ or C$_6$F$_6$ liquid scintillators of typical 1 l
volume, which allow the detection of capture events with
$\approx$20\% efficiency but require an external weighting
function for obtaining the linear relationship between
$\gamma$-ray energy and efficiency \cite{MaG67b}. These detectors,
which are currently in use at electron linear accelerators
\cite{KSW96,BCM97}, do not provide sufficient gamma-ray energy
information to differentiate capture on the sample from capture of
scattered neutrons on nearby materials. Hence, corrections for the
various backgrounds discussed below are much more difficult to
determine.

\subsection{Choice of Scintillator}

The main problem in using 4$\pi$ arrays results from their much larger
detector volume and from the fact that the chemical composition of the
inorganic scintillator may include isotopes with relatively large
cross sections for neutron capture. This makes them much more sensitive to
neutrons scattered
by the sample. Particularly for keV neutron energies, where scattering
in the sample dominates over capture, subsequent capture of the scattered
neutrons in the scintillator can increase the overall
systematic uncertainty.  These effects reduce the number of suitable
scintillator materials. For example, NaI and CsI have to be excluded from
this application because of the large iodine (n,$\gamma$) cross section.
Several suitable scintillators are listed in Table\,\ref{tab_scint}, all
being composed of low cross section materials.

\begin{table}
\caption{Characteristics of some scintillator materials}
   \label{tab_scint}
   \begin{tabular}{lcccc}
 Scintillator           & Density    & Decay time & Wavelength & Photons per MeV \\
                        & (g/cm$^3$) & (ns)       & (nm)       &                 \\
    \hline
 BaF$_2$                & 4.88       & 0.6; 630   & 220; 310   & 1800; 10000     \\
 Bi$_4$Ge$_3$O$_{12}$ (BGO) & 7.13       & 60; 300    & 480        & 700; 7500   \\
 CeF$_3$                & 6.16       & 3; 27      & 300; 340   & 200; 4300       \\
 C$_6$F$_6$             & 1.61       & 3.3        & 430        & $\approx$10000  \\
 \end{tabular}
 \end{table}

The choice of BaF$_2$ for the Karlsruhe 4$\pi$ array \cite{WGK90a} was
based on the fact that it exhibits similar sensitivity to scattered
neutrons compared to CeF$_3$ and bismuth germanate (BGO) but has the
advantage of better time and energy resolution. The responses of the
scintillators listed in Table\,\ref{tab_scint} to scattered neutrons
are compared in Table\,\ref{tab_sens}.  (See Section V.C.8 for details of
the calculation.) Here we assume scintillator volumes of
equal $\gamma$-efficiency, a capture sample of gold, and a well
collimated neutron beam. One finds that the organic C$_6$F$_6$
scintillator has by far the smallest sensitivity for scattered neutrons,
despite the large thickness required to match the efficiency of the
high-Z materials in the inorganic scintillators. With BaF$_2$, comparable
values could only be reached using isotopically pure $^{138}$Ba due to
the very small (n,$\gamma$) cross section of this neutron magic
nucleus.

\begin{table}
\caption{Neutron sensitivity of suited scintillators in different energy regions assuming a bare gold sample in a well collimated beam and neglecting the delay
between capture of neutrons in the sample and capture of scattered neutrons in the 
scintillator.}
   \label{tab_sens}
   \begin{tabular}{cccccc}
 Scintillator & Thickness & \multicolumn{4}{c}{Capture of scattered neutrons/true 
 captures in sample}\\
              & (cm)      & 0.1 to 1 keV    & 1 to 10 keV     & 10 to 100 keV   & 
	      0.1 to 1 MeV \\
    \hline
 BaF$_2$      & 15        & 0.52$\pm$0.02   & 1.09$\pm$0.04   & 1.34$\pm$0.10   & 
3.19$\pm$0.60 \\
 Bi$_4$Ge$_3$O$_{12}$ (BGO)
              & 10        & 0.35$\pm$0.02   & 0.84$\pm$0.05   & 1.24$\pm$0.14   & 
2.22$\pm$0.61 \\
 CeF$_3$      & 13        & 0.038$\pm$0.004 & 0.12$\pm$0.01   & 0.75$\pm$0.10   & 
1.89$\pm$0.48 \\
 C$_6$F$_6$     & 60      & 0.035$\pm$0.004 & 0.018$\pm$0.005 & 0.036$\pm$0.014 & 
1.28$\pm$0.37 \\
 \end{tabular}
\end{table}

To minimize the neutron sensitivity, all considered scintillators are
composed of materials with small capture cross sections and hence large
scattering/capture ratios. Accordingly,
most scattered neutrons diffuse out of the 4$\pi$ array without being
captured. Typical diffusion times being of the order of 3 $\mu$s
imply that the related captures would appear as prompt background in
experiments using long neutron flight paths, whereas most of the capture can be
discriminated via time-of-flight in experiments carried out with very short
flight paths such as those used with low energy, monoenergetic sources.

\subsection{Detector details}

There are various possibilities to cover the full solid angle with an
arrangement of closely packed crystals. In order to facilitate the
interpretation of event multiplicities, the crystals should be shaped
such that they cover equal solid angles. Such geometries are known to
correspond to fullerene-type structures consisting of a few
configurations with certain, fixed number of elements \cite{HSD79}.
In this work two geometries with 42 and 162 crystals respectively were
simulated.

The 42-element geometry using 30 hexagonal and 12 pentagonal
crystals was adopted for the Karlsruhe 4$\pi$ BaF$_2$ detector
\cite{WGK90a}. In this approach, the
crystals form a closed sphere with an inner diameter of 20 cm and
a thickness of 15 cm, which is sufficient to detect gamma rays
of a few MeV energy, the minimum of the absorption coefficient, 
with $\ge$90\% efficiency. 
The geometry of this setup was modeled in detail,
including the light reflector (Teflon) and the aluminum cladding
of the individual detector modules. These simulations could be
verified by comparison with the experimental performance of the
detector (Sec.\,III).

\begin{figure}
\includegraphics[width=.9\textwidth]{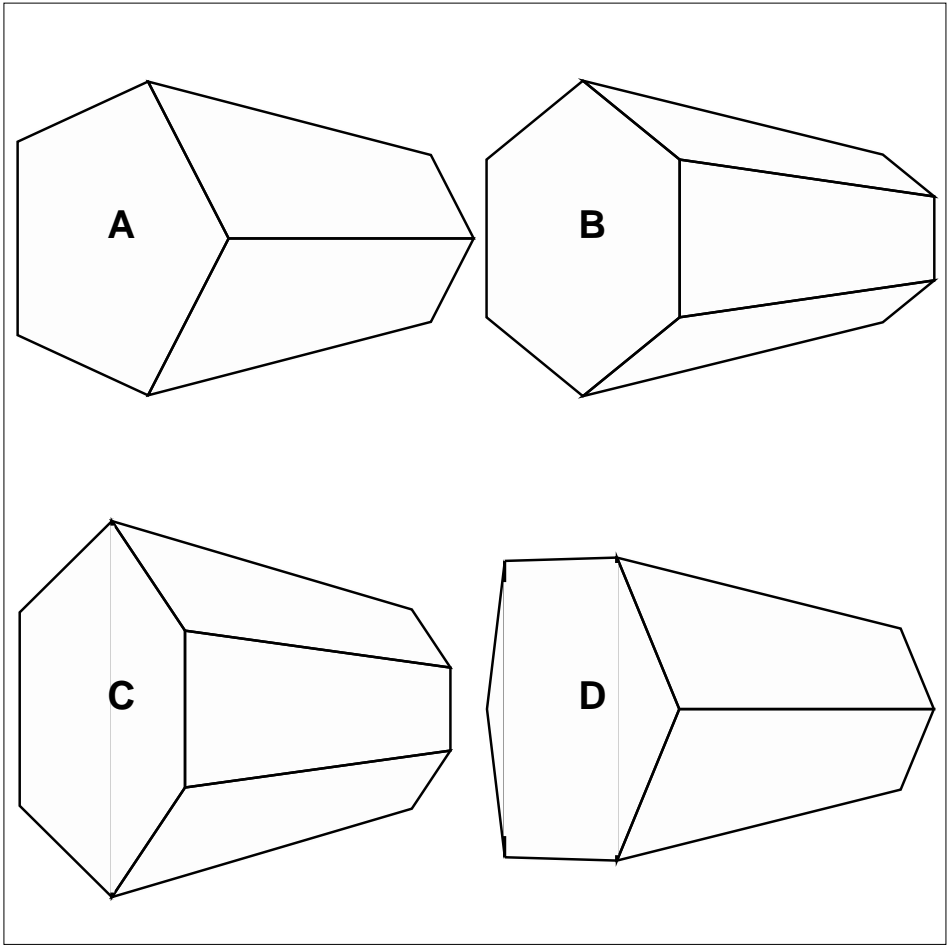}
 \caption{ A 4$\pi$ array with 42 modules
 can be built from 12 pentagons and 30 regular hexagons (see
 Ref.\,\protect\cite{WGK90a}). The
 162-module design is also based on 12 regular pentagons of type A (left, top) 
 but requires three different shapes for the hexagons, 30 regular crystals of type B, 
 60 of type C, and 60 of type D (right, bottom).
 \label{fig_arrays}}
 \end{figure}

The setup with 162 elements and therefore higher granularity is
favored for experiments where high count rates or higher
multiplicities are expected. This more complex geometry requires
four different crystal shapes to cover the
sphere uniformly, as shown in Fig.\,\ref{fig_arrays}. A
full model was also constructed for this geometry including all
details used in the 42-fold scintillator array.

In order to find an
optimized solution with respect to several design parameters,
simulations were carried out for different crystal thicknesses and for
different inner diameters of the 4$\pi$ detector. With a shell thickness of
15 cm a total efficiency of $\ge$90\% for $\gamma$-rays in the
relevant energy range could be
achieved. Increasing the thickness of the crystals yields a higher
efficiency but the larger detector volume implies more severe
background due to capture of sample-scattered neutrons. For this reason
and for controlling the overall costs, the inner diameter of the
detector should be kept as small as possible though it was found that
this background can be substantially reduced by means of a $^6$LiH
liner inside the detector sphere. In summary, an inner diameter of 20 cm
and a crystal thickness of 15 cm was found to be a good compromise
between detector performance and costs.

\section{GEANT simulations}
\noindent

The performance of the various design options was simulated with the
GEANT detector description and simulation tool
from CERN \cite{GEA93}. GEANT tracks photons, electrons and hadrons,
which in this case are only neutrons and protons. In this work,
the GHEISHA module was used for tracking neutrons with energies
above 200 keV, while the GCALOR \cite{ZeG94} interface with the MICAP package
was preferred for neutron energies below 200 keV. A 10 keV cut-off
energy was chosen for all particles except for neutrons, where this
limit was defined at 10 meV. Particles falling below these thresholds
are assumed to deposit their residual kinetic energy without further
interaction. In general, GCALOR considers pointwise cross section
data from the ENDF/B-VI evaluation \cite{NND96}. Additionally,
theoretical $\gamma$-cascades for neutron capture events had to be
used for the barium isotopes and for gold \cite{UhK93,WVK95b} since
these photon data for neutron capture events were not available
from this library.

The calculations account only for the energy deposited in the
scintillators and neglect photon losses and the limited
photo-efficiency of the photomultiplier tubes (PMs). In the
simulation these effects were considered by folding the calculated
response with the experimentally determined energy resolution of
the Karlsruhe 4$\pi$ BaF$_2$ detector, which uses both the short
and long decay-time components of the light output, which together
give about 11,800 photons per MeV. The effects of photon yields of the
other scintillators and for using just one component of the BaF$_2$
scintillation
light is to change the resolution approximately by the photon statistics,
$\approx$ 1/$\sqrt{N}$. For C$_6$F$_6$ detectors, where the size of the
scintillator
must be very large in order to absorb the entire $\gamma$-ray energy, further
losses of light are expected to reduce the resolution.

\begin{figure}
\includegraphics[width=.9\textwidth]{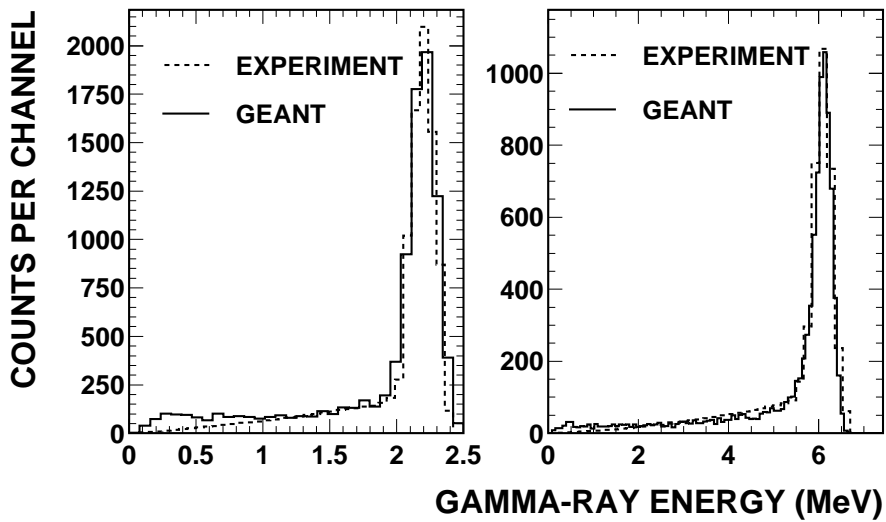}
\caption{Measured and simulated response of the Karlsruhe array of 42
BaF$_2$ modules to monoenergetic $\gamma$-rays. 
\label{fig_mono}}
 \end{figure}

Since GEANT is a simulation tool developed for tracking high energy
particles, the treatment of low-energy interactions 
may be uncertain by 20\% \cite{GEA93}. This
problem was checked by comparing GEANT simulations for monoenergetic
$\gamma$-rays with the experimental spectra measured with the
Karlsruhe array.  In this experiment, the proton beam of the 
Van de Graaff accelerator was directed onto
thin $^{26}$Mg, $^{30}$Si, and $^{34}$S targets in the center
of the detector. These targets exhibit (p,$\gamma$) resonances,
which are known to produce pure two-step cascades. By replacing the
BaF$_2$ crystal at zero degrees by a HPGe detector and by
gating the response of the
4$\pi$ detector by requiring a full-energy signal in the Ge detector,
the response of the Karlsruhe array to monoenergetic $\gamma$-rays
could be measured for 22 energies from 0.843 MeV to 8.392 MeV
\cite{Web93}. The experimental spectra for 2.209 MeV and 6.146 MeV
$\gamma$-rays are compared in Fig.\,\ref{fig_mono} to the corresponding
GEANT simulations. Obviously, the measured spectra are
reproduced very well in the simulations, except for small deviations 
at lower energies due to the effect of differences in the electronic 
thresholds.

\section{The Karlsruhe 42-Detector 4$\pi$ Array}

\subsection{The facility}

The Karlsruhe 4$\pi$ array is used for the determination of
(n,$\gamma$) cross sections in the astrophysically relevant
neutron energy range from 3 to 220 keV. The present simulation of
this detector concentrated on the measured signature
from the prompt capture $\gamma$-ray cascades and on the
background due to sample-scattered neutrons which  may be captured
in materials of the scintillator (BaF$_2$, Al, Teflon, etc.). With
the Karlsruhe neutron source, time-of-flight information can be used to
separate these two processes. The most important features for
characterizing these two components are the respective sum energy
signals and the corresponding multiplicity distributions.  Because
the Karlsruhe detector is so well characterized, its performance
can serve as a test and validation of the GEANT simulations.

\begin{figure}
\includegraphics[width=.9\textwidth]{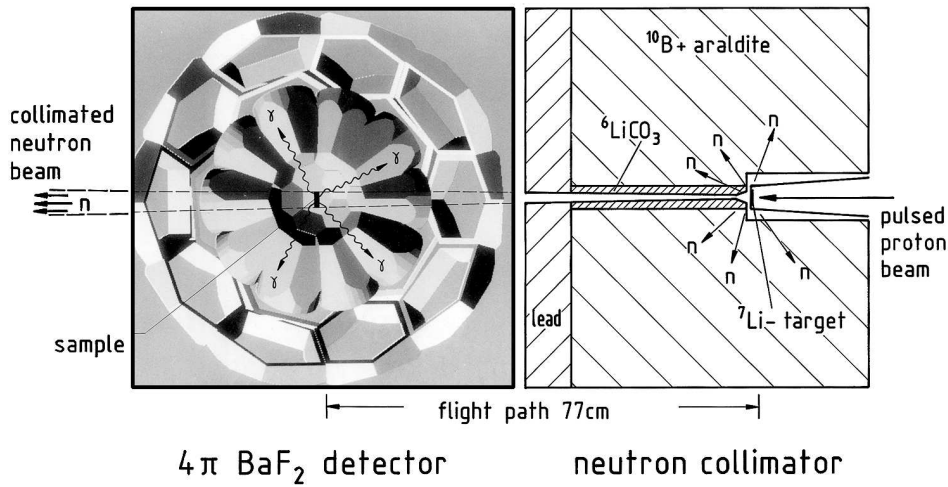}
\caption{The Karlsruhe setup for (n,$\gamma$) studies with the 42-module 
BaF$_2$ array. 
 \label{fig_4pika}}
 \end{figure}

The essential features of the experimental setup are sketched
in Fig.\,\ref{fig_4pika}. Neutrons with a continuous energy distribution are
produced via
the $^7$Li(p,n)$^7$Be reaction by bombarding thin metallic lithium targets
with the pulsed proton beam from a Van de Graaff accelerator (repetition
rate 250 kHz, beam energy 1.9 to 2 MeV, average beam current 2 $\mu$A).
The collimated neutron beam hits the sample in the center of the detector
array after a flight path of $\approx$80 cm.

\subsection{Response to gamma rays from capture on the sample}

The simulations required information on
the original capture cascades, which were obtained either from the ENDF/B
library that is part of the GCALOR software or from detailed theoretical
calculations. Throughout this paper, the simulations refer to metallic
gold samples. Since such samples are routinely used for neutron flux
determination in all measurements with the 4$\pi$ detector, there are
ample data available for comparison. In simulating the background from
scattered neutrons (Section III.C), capture events in the various barium
isotopes as well as in fluorine and reflector materials were considered.

\begin{figure}
\includegraphics[width=.9\textwidth]{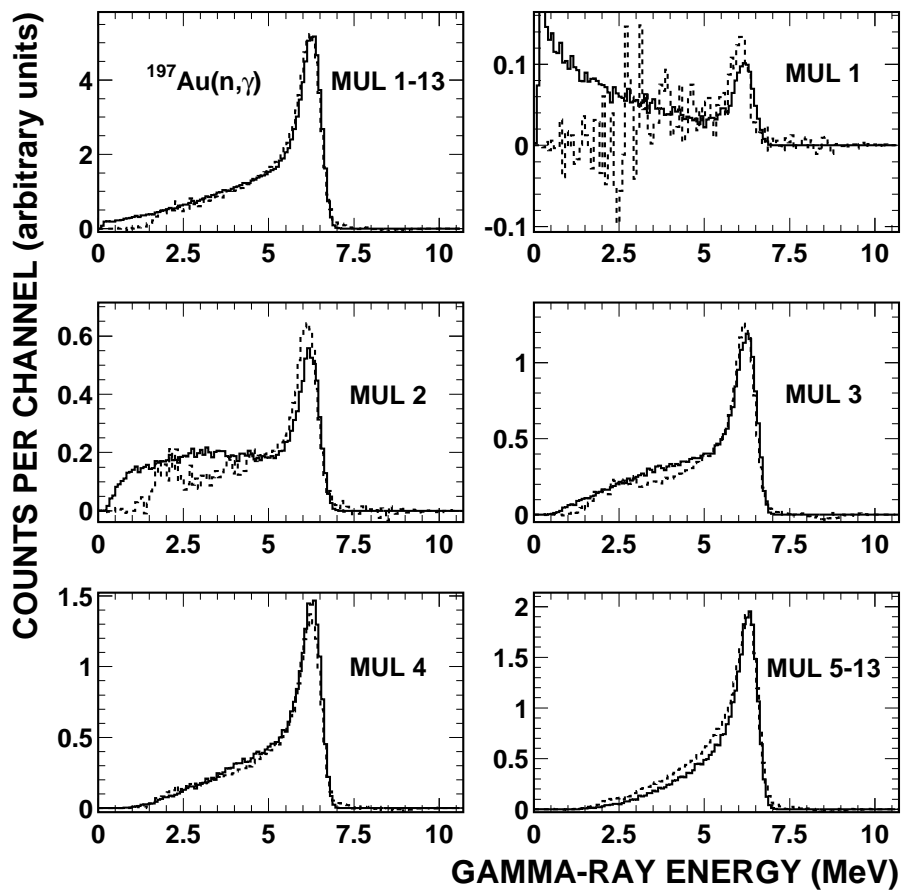}
\caption{Measured (dashed curve) and simulated (solid curve) response of the 
Karlsruhe 4$\pi$ detector to neutron captures in $^{197}$Au. The spectrum
averaged over all events is shown in the left panel in the top row. The spectra for 
events with different multiplicities can also be well reproduced.
\label{fig_auka}}
 \end{figure}

The first example is presented in Fig.\,\ref{fig_auka}, showing the comparison
between the experimentally measured sum-energy spectra of neutron
captures in the gold sample \cite{WVK92} and the corresponding
simulation. The
spectra show two components: capture cascades which are completely
detected and appear as a line at the binding energy of the captured
neutron, and events where part of the energy of the cascade $\gamma$-rays
escapes detection resulting in a tail towards lower energies.  Both
components are reproduced very well by the simulation.

It is quite remarkable that the successful agreement between experiment
and simulation holds not only for the average of all events but rather well
also for the different detector-hit multiplicities.  The various
multiplicity cuts contain only one normalization constant, the total
number of events.  In achieving
this result it was also important to consider the internal conversion 
of low-energy $\gamma$-transitions in the theoretical cascades.
Particularly for the gold sample, this effect caused a significant
broadening of the peak toward lower total detected energy. 

\subsection{Response to scattered neutrons}

\begin{figure}
\includegraphics[width=.9\textwidth]{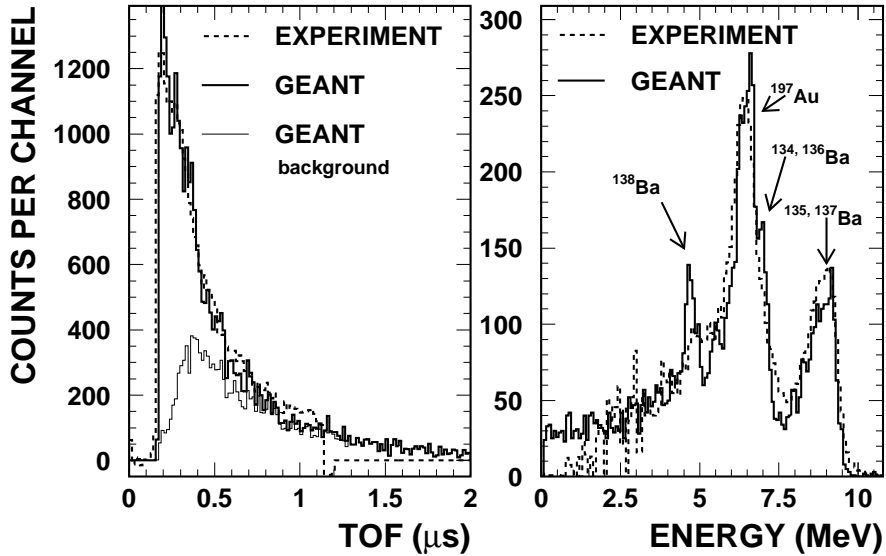}
\caption{Simulation of background due to capture of scattered neutrons in the 
Karlsruhe array. Left: Comparison for a TOF spectrum. 
Right: The $\gamma$-ray spectrum of the same events 
showing that different background reactions in the even and odd Ba isotopes
can be distinguished. 
\label{fig_tofka}}
\end{figure}

In a further step, the background from sample-scattered neutrons was
simulated for the experimental TOF spectra. The left part of
Fig.\,\ref{fig_tofka}
shows a typical experimental TOF spectrum, the edge at short times
reflecting the maximum neutron energy of 130 keV. The cut-off at 1.15
$\mu$s corresponds to the software window for suppressing the neutron
energy range below 2.5 keV, which is dominated by the background from
scattered neutrons. The same data are plotted in the right part of
Fig.\,\ref{fig_tofka} but projected on the $\gamma$-energy axis, illustrating
the
advantage of a calorimetric measurement with good energy resolution.
This spectrum exhibits a number of peaks corresponding to capture
events in different isotopes in the detector, thus allowing unambiguous
background assignments.

Again, both spectrum types in Fig.\,\ref{fig_tofka} are fitted very well by the
simulations. The quality of the simulations can be appreciated by the
significance of the small differences in the right spectrum, which
were identified as being due to errors in the original barium cross
sections used in the ENDF/B-VI library.  Recently, this discrepancy could be
explained on the basis of improved Ba cross section data
\cite{Kae98}.

In summary, these examples studied in this section indicate that the
rather complex geometry of a 4$\pi$ detector of high granularity can
be successfully described. The successful simulation of the detector
response to monoenergetic $\gamma$-rays as well as to the experimental situation
of
an actual measurement makes the GEANT software a reliable tool for
studying and optimizing this type of detectors in a variety of
applications.

\section{GEANT simulations for setup at spallation sources}
\subsection{General Considerations}

Neutron capture experiments at a spallation, white neutron source
face challenges not encountered at monoenergetic or
quasi-monoenergetic sources. Firstly, because of the long flight
path betweeen the source and sample and because of the presence of
higher energy neutrons, time-of-flight cannot be used to
differentiate neutrons captured in the sample from neutrons
scattered by the sample and subsequently captured in the BaF$_2$
scintillator.  Moderation and absorption of the scattered neutrons
is an approach to reduce this background.  Secondly, collimation
of the higher energy neutrons is more difficult and the ambient
background is likely to be higher.  Because these effects are
specific to individual flight paths at given facilities, only the
effect of an assumed beam halo is discussed here. Thirdly, the
duty factor of monoenergetic sources is high, sometimes nearly
100\%. Spallation sources typically have repetition rates of  20
to 30 Hz \cite{MiW90} and the new source at CERN is planned to run
between 0.4 and 0.07 Hz \cite{AAA99}. Longer running times
are therefore needed at spallation neutron sources if the number
of events per beam burst is the same. However, the instantaneous
rate at spallation sources is much higher, which enables the use
of small samples of very radioactive materials. The
high instantaneous rate, however, requires that new techniques be
developed to handle the large amounts of data.

Unless otherwise specified, all following simulations refer to a
160 crystal BaF$_2$ array (162 elements with vacant entrance and 
exit positions to accommodate the beam tube), a 1/E approximation to the
neutron spectrum expected for the LANSCE TOF facility at 
20 m flight path \cite{HRK99}, and a gold sample in the center
of the array. Note that gold is a relatively forgiving sample as
far as the scattering background is concerned, the
scattering/capture ratio ranging between 10 and 100 at keV neutron
energies.  For other possible samples, the ratio can range to well
over 1000 and the discrimination against capture of scattered
neutrons by the detector material is then more challenging.

\subsection{Spallation Neutron Spectra}

The neutron spectra from spallation neutron sources have been discussed
previously \cite{MiW90,CHK92}. For the present considerations, the neutron
spectrum is that emitted from a moderator near the primary neutron
production site to enhance the neutron flux below 100 keV. At LANSCE
this spectrum has been measured to be E$^{-0.948}$, which is very close
to 1/E, up to 100 keV \cite{Koe99,HRK99}. The shape of the spectrum at
higher energies has been calculated and needs to be measured. For the LANSCE
simulations reported
here, we assume that the spectrum continues as 1/E at all energies. The
CERN facility is still in the planning stages and for its neutron spectrum
we rely on calculations \cite{AAA99}. Note that there is an increase over
the 1/E shape at neutron energies between 10 keV and 1 MeV.

\subsection{Response to scattered neutrons}
\subsubsection{Baseline calculation}

As in the Karlsruhe setup, the capture detector at a spallation neutron source
will be sensitive to neutrons scattered from the sample.  With the 20 meter
flight path and the 1/E neutron spectrum at LANSCE, pulse height spectra
were calculated for flight times corresponding to neutrons in the range 0.1 to
1eV, 1 eV to 10 eV, and so forth.  Pulses due to true capture events on the gold
sample were tallied and compared with pulses due to scattered neutrons captured
in the detector.  The results, given in Fig. 6, show that the neutron
sensitivity of the detector becomes important for times corresponding to neutron
energies above 1 keV.  Above 100 keV, the detector response is dominated by
events due to these scattered neutrons.

\begin{figure}
\includegraphics[width=.9\textwidth]{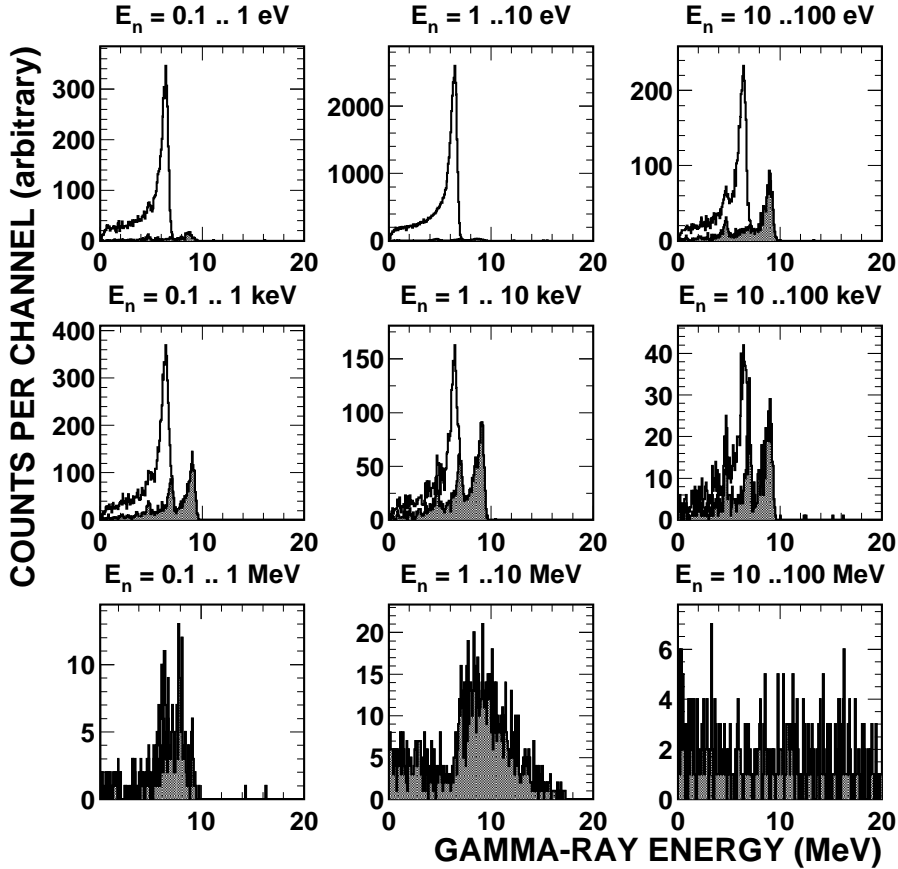}
\caption{Simulated $\gamma$-ray spectra (solid line) for capture events 
expected at the neutron spectrum of a spallation source. The respective
 background due to scattered neutrons (shaded area) is indicated for different
 time-of-flight bins. 
\label{fig_ensim}}
\end{figure}

In the following subsections, we discuss design parameters that affect the
ability to differentiate true capture events from those due to scattered
neutrons.

\subsubsection{Flight path lengths}

Apart from the good resolution in time-of-flight, the 
200 m flight path proposed at CERN \cite{AAA99} eases also the task
of discriminating between true capture events and those related to
capture of neutrons scattered from the sample.  The reason is that
at CERN there will be more separation in time between neutrons of
different energies.  The scattering of higher energy neutrons from
the sample (where the scattering-to-capture ratio is very high)
will thus be more easily separated from the desired capture
events from lower energy neutrons, an effect which can be important
in case of resonance-dominated cross sections.

\subsubsection{Effect of using only the fast component of BaF$_2$ scintillation}

\begin{figure}
\includegraphics[width=.9\textwidth]{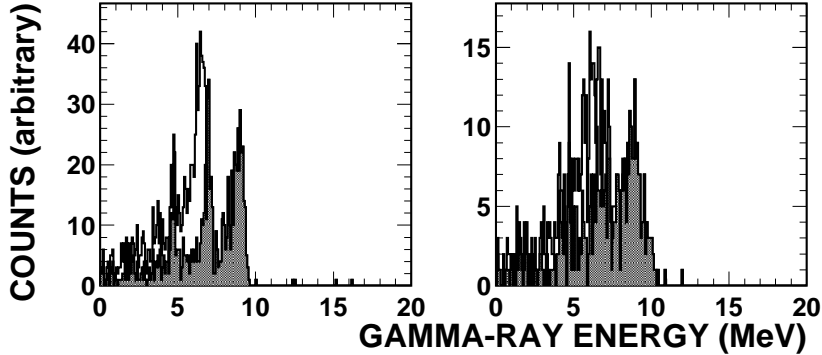}
\caption{Comparison between the pulse height spectra for captured and scattered neutrons 
with the Karlsruhe energy resolution (left) and the expected resolution if only the fast 
component of the BaF$_2$ scintillators were used (right). The neutron energy range was 10 
keV to 100 keV. Response to scattered neutrons is indicated by the shaded spectrum. 
Note the importance of resolution in $\gamma$-ray energy.
\label{fig_simfast}}
\end{figure}

In order to avoid pile-up effects at
the higher neutron fluxes at spallation sources, it might be
necessary to integrate only the fast component of the scintillator
light, which contributes about 15\% to the total light output.
Correspondingly one would expect a reduction in energy resolution
by a factor of $\sqrt{7}$. The effect of such a reduced energy
resolution with respect to the background from capture in barium
is shown in Fig.\,\ref{fig_simfast}.  For some samples, especially those with Q-
values for neutron capture similar to those for barium isotopes, the degradation
in resolution makes more difficult the separation of signal from this background
of scattered neutrons captured in the scintillator.

\subsubsection{Effects of sample backings }

Possible backing materials for target preparation were
investigated with respect to their background contributions due to
neutron capture and scattering. Disk samples of 1 cm diameter were
positioned in the center of the detector, and the thickness
of the backing was assumed to equal the thickness of the gold
sample. The results obtained for carbon and
beryllium backings are summarized in Table\,\ref{tab_back}, which
compares the background due to scattering in the backing and the
number of true capture events in the gold sample for different
neutron energy intervals. The scattering contribution from the
gold layer is listed separately for easier comparison.

\begin{table}
 \caption{Background contributions from sample backings for a gold layer with the same thickness and diameter as the backing.}
   \label{tab_back}
   \begin{tabular}{ccccc}
 Backing & \multicolumn{4}{c}{Backing-related events (scatter followed
 by capture}\\
    & \multicolumn{4}{c}{in the scintillators)/true captures in a gold sample}\\
              & 0.1 to 1 keV    & 1 to 10 keV     & 10 to 100 keV   & 0.1 to 1 MeV 
\\
    \hline
 Carbon    & 0.33$\pm$0.02 & 0.50$\pm$0.04 & 1.13$\pm$0.13 & 1.21$\pm$0.36 \\
 Beryllium & 0.51$\pm$0.02 & 0.80$\pm$0.05 & 1.77$\pm$0.18 & 1.43$\pm$0.40 \\
 Gold      & 0.52$\pm$0.02 & 1.09$\pm$0.04 & 1.34$\pm$0.10 & 3.19$\pm$0.60 \\
 \end{tabular}
 \end{table}

The simulations show, not surprisingly, that the background is
dominated by neutron scattering from the backing unless the sample
thickness is considerably larger than that of the backing.  For
example a 25 $\mu$m beryllium substrate with a 0.5 $\mu$m gold
target (1 mg/cm$^2$) would have a ratio of 90 scattered neutron 
events to 1 capture event in the 10 to 100 keV neutron energy region.
Therefore sample diameters smaller than 1 cm are preferable.
Titanium may be preferable as a backing material because of its
high tensile strength and low neutron-capture cross section.  
Titanium was not evaluated in these
simulations because the scattering cross sections were not in the
GCALOR database.

\subsubsection{Effect of beam pipe }

The neutron beam from the spallation source will be in vacuum. Unlike the
situation at Karlsruhe where the source is relatively close to the sample, the
vacuum is necessary to eliminate neutron scattering from 20 to 200 meters of
air, which would seriously degrade the beam quality. Furthermore, near the
sample each cubic centimeter of air has approximately the same mass as the
sample and therefore would contribute significantly to the effects of scattered
neutrons. Gamma rays from capture on
the sample as well as neutrons scattered from the sample will interact with the
beam pipe.  Selection of material for the beam pipe is critical. For
aluminum, which has a small neutron capture cross section as well as a
fairly low atomic number, most of the gamma rays will pass. Choosing a
beam pipe with an outer diameter of 3 cm, some
degradation in the total energy signal is observed when the pipe has a 3 mm wall
thickness (see Fig. 15 in Ref.\,\cite{HRK99}).

\subsubsection{Neutron absorbers}

\begin{figure}
\includegraphics[width=.9\textwidth]{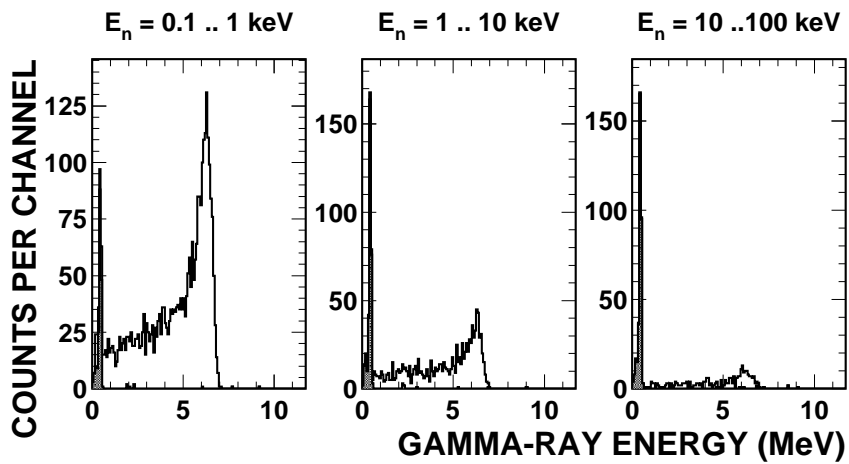}
\caption{The $\gamma$-ray spectra obtained with a spherical $^6$LiH absorber 
surrounding the sample and with thin $^{10}$B layers between the modules. 
The background from scattered neutrons (shaded regoins) is almost eliminated. 
The line at low energies stems from the 478 keV transition following the
$^{10}$B(n,$\alpha$) reaction.
\label{fig_absorb}}
\end{figure}

In order to reduce the background due to scattered neutrons, the
possibility of absorbing scattered neutrons by suitable materials
was investigated. For comparison with the unshielded configuration,
three simulations are presented.  First 0.8 mm thick layers of
$^{10}$B were inserted between the various BaF$_2$ crystals.
Secondly, an inner 8 cm thick moderating neutron absorber of $^6$LiH 
was placed just inside the inner radius of the BaF$_2$-array. Finally a 
combination of both approaches was simulated (see Fig.\,\ref{fig_absorb}). 
In total, a background reduction by a factor 100 could be achieved in
the astrophysically relevant region between 10 and 100 keV.  The
combination of the two absorbers was more effective than one might
expect based solely on the simulations of the single absorbers.

\begin{figure}
\includegraphics[width=.9\textwidth]{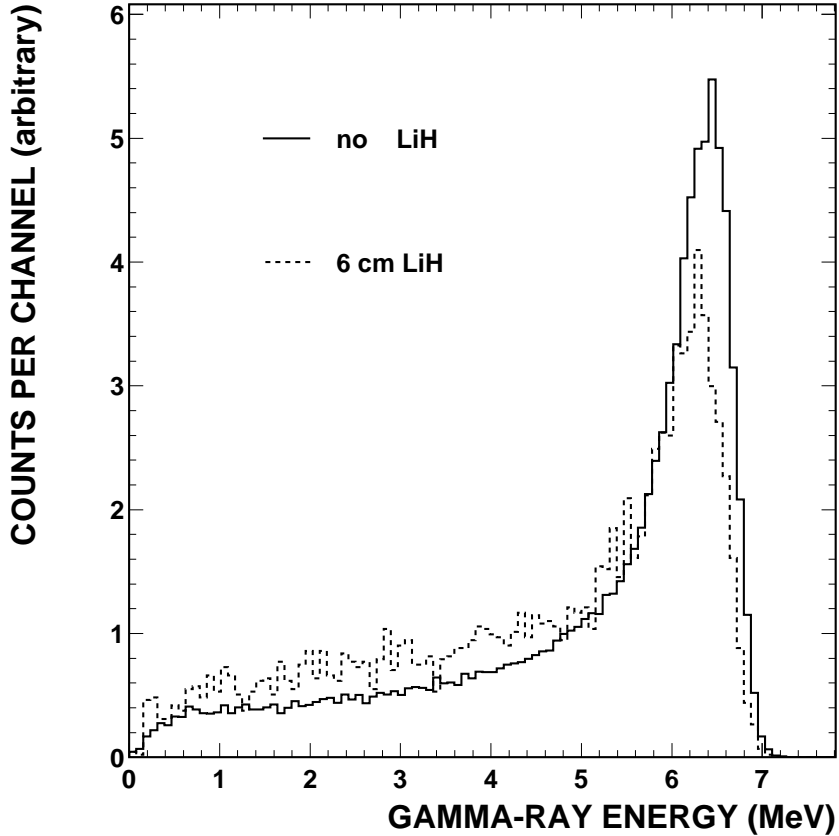}
\caption{Influence of a 6 cm thick $^6$LiH moderator/absorber layer 
on the shape of the $^{197}$Au$(n,\gamma)$ spectrum.
\label{fig_abres}}
\end{figure}

It is apparent that including a $^6$LiH absorber
may cause some degradation of the capture $\gamma$-ray
spectrum. This effect is illustrated in Fig.\,\ref{fig_abres} for
different moderator/absorber thickness. A thick lithium
hydride shell degrades the energy resolution of the detector and
increases the low energy tail.  Obviously the effect on the $\gamma$-rays
is relatively small, however, since LiH consists of low-Z elements.

The use of neutron absorbers as described above reduces the ratio
of scattered neutron events to capture events in the 10 to 100 keV
neutron energy region from 90 to 0.9 for a 1 mg gold target on a
25 $\mu$m beryllium substrate.  This is sufficient to allow the
further subtraction of scattered neutron events based on the
energy spectrum.  Greater improvement may be possible by using
thinner target substrates such as titanium foils.

\subsubsection{Detector granularity}

\begin{figure}
\includegraphics[width=.9\textwidth]{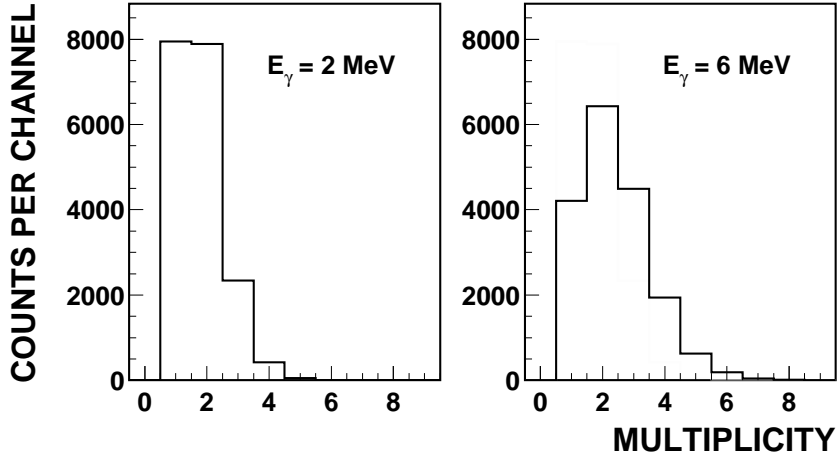}
\caption{Multiplicity distribution of detectors hit for monoenergetic $\gamma$-rays
 observed with the 162 module array. 
\label{fig_mult}}
\end{figure}

The high granularity of a 162-element array introduces the
possibility of analyzing the pattern of detectors that are hit in
order to help discriminate between $\gamma$-ray and scattered-neutron
induced events.  This is not an easy problem, since gammas above
0.5 MeV interact primarily by Compton scatter and, at higher energies,
by pair production, producing electron-gamma showers in the array.  Thus,
even single gammas can cause several crystals to fire.  This is illustrated 
in Fig.\,\ref{fig_mult}, which shows the calculated multiplicity, or number of
crystals that fire, for $\gamma$-rays of 2 and 6 MeV emitted at random angles 
from the center target position in the BaF$_2$ ball. Gamma rays of other 
energies were also investigated in the range from 1 to 9 MeV. For almost all 
energies, multiplicity 2 is the most frequent.  The average multiplicity (M) 
was found to be a linear function of the $\gamma$-energy (E$_\gamma$) in MeV, 
${M = 0.154 \cdot E_\gamma + 1.44}$.

\begin{figure}
\includegraphics[width=.9\textwidth]{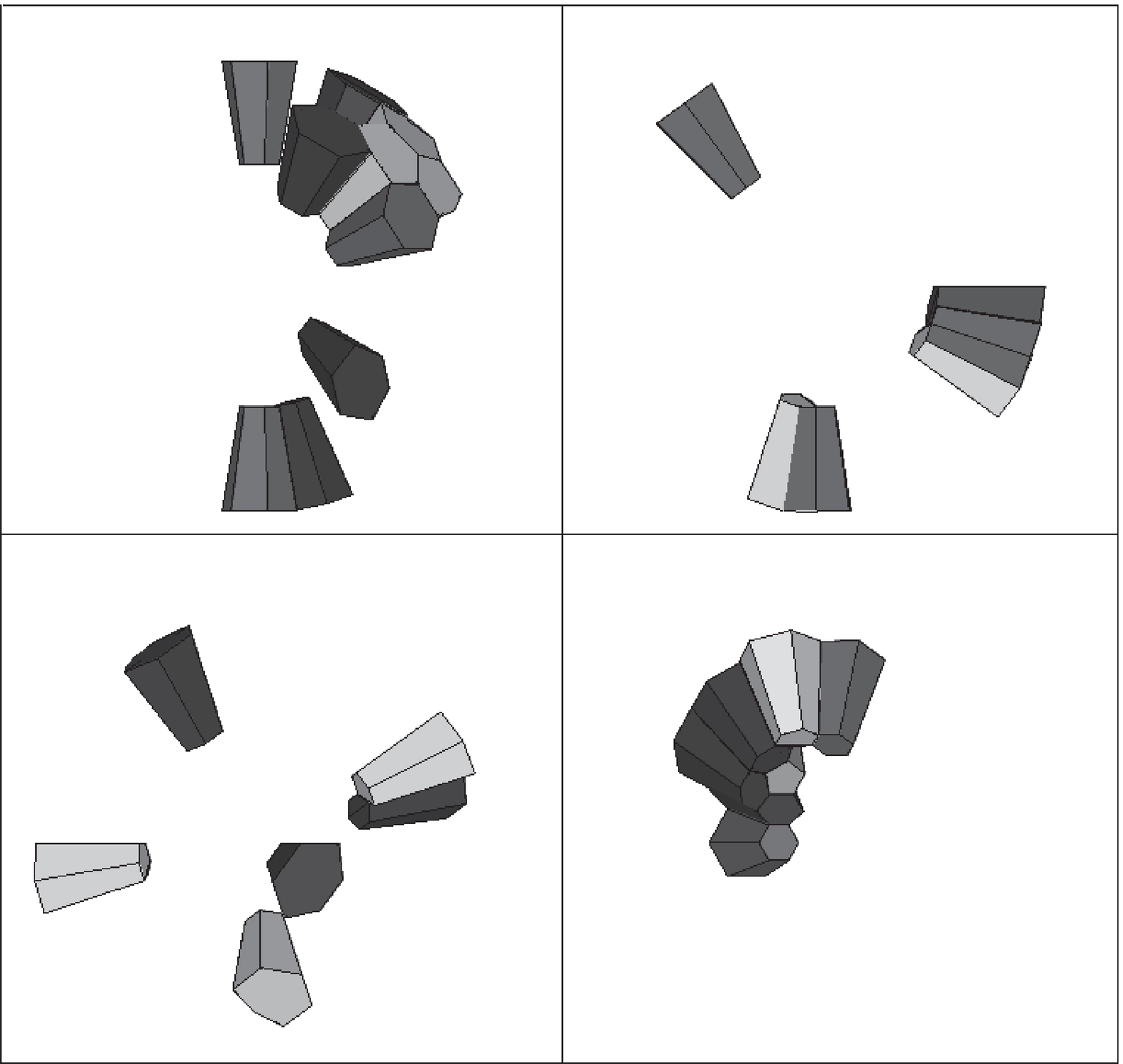}
\caption{Sample hit patterns for monoenergetic $\gamma$-rays of 6 MeV in the 162
 module array. Shown are those modules which recieve an energy deposit  in excess 
of 50 keV. Obviously, hit patterns with higher multiplicities are not restricted to
adjacent crystals.
\label{fig_hit}}
\end{figure}

Sample hit patterns for multiplicities greater than two from
monoenergetic 6 MeV gammas are shown in Fig.\,\ref{fig_hit}.  The figure
illustrates that the hit crystals do not have to be adjacent.

In an actual capture event, whether in the target or induced by a
scattered neutron in a crystal, several gamma rays can be emitted
resulting in a multiplicity distribution with the most probable
multiplicity greater than 2.  The simulations showed that there
was no usable difference in the multiplicity distribution between
capture and scatter events.   However, captures that occur in a
specific crystal, due to scattered neutrons, will produce several
gammas which will most likely fire adjacent crystals.  Multiple
gammas produced by capture events in the target will most likely
hit crystals in different regions of the array.  Thus scatter
events are expected to produce a single large cluster of hits,
while capture events will produce multiple smaller clusters.

Clustering was analyzed in the simulation by tallying the number
of hit clusters that occur for each event, the number of crystals
that fire in the largest cluster, and the total energy deposited
in the array divided by the number of clusters.  A cluster was
defined as a set of neighboring crystals that are all hit, and two
crystals are defined as neighbors when the angle between two rays
drawn from the center of the array to the center of each crystal
is less than 20 degrees.  As expected, the simulation showed that events 
due to scattered neutrons have fewer clusters, but more crystals are hit
in the largest cluster.  Fifty percent of events caused by
scattered neutrons formed only one cluster.

\begin{figure}
\includegraphics[width=.9\textwidth]{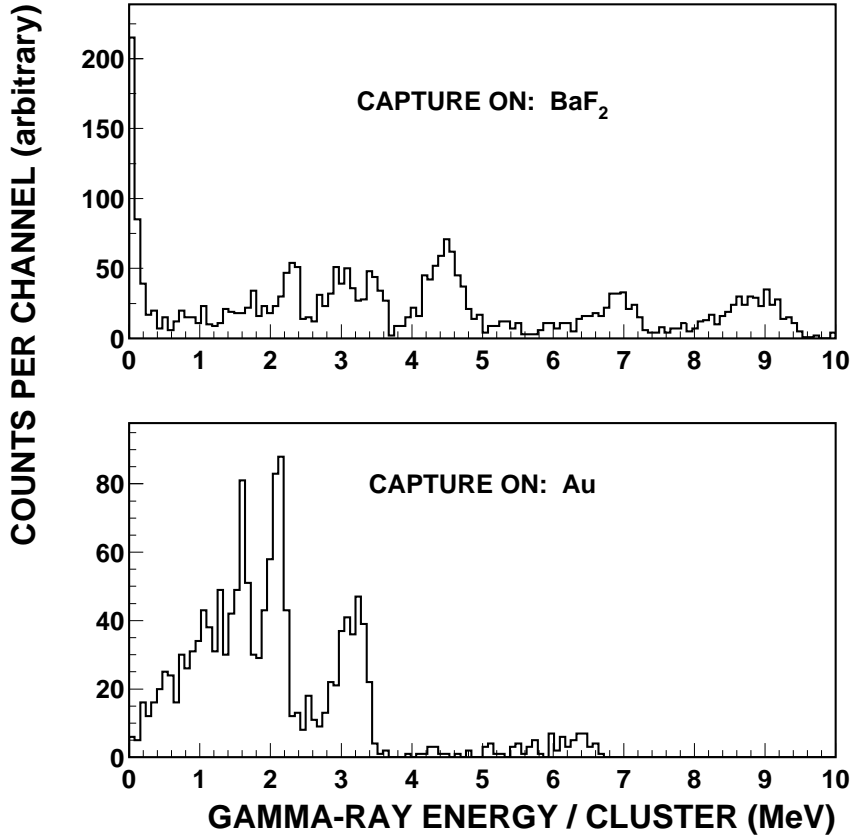}
\caption{Gamma-ray energy per cluster for neutron scatter events resulting in capture in the 
BaF$_2$ array (top) and for capture events in a gold sample (bottom). This difference provides
a possible means of discriminating between capture and scatter events.
\label{fig_hitrec}}
\end{figure}

Fig.\,\ref{fig_hitrec} shows the distribution of the energy per cluster.  
The average energy extends higher in events caused by scattered neutrons.
This is obvious since the total energy released as gammas is
roughly the same for all reasonably heavy nuclei, but it can be
all deposited in one cluster in a scatter event, while it is
distributed over several clusters in a capture event.  The best
separation between events due to capture in the target and events
due to scattered neutrons can be made by requiring that more than
one cluster be formed, and that the average energy per cluster be less
than 3.8 MeV.  With these cuts, the ratio of scatter to capture
events can be reduced by a factor of 3.

It should be noted that the simulations were made with Au as a
target.  These results would be different if the $\gamma$-spectrum of
the actual target were significantly different than that of Au.
These results also relied on the energy resolution obtained by
using both the fast and slow components of the BaF$_2$ crystal.  If
only the fast component were used, energy cuts may not be possible
due to the poorer energy resolution (see below).

\subsubsection{Other detector materials}

Various scintillator materials can be used for $\gamma$-ray
spectroscopy. In this section different scintillator materials are
investigated with respect to their neutron sensitivity. Other
properties, e.g. decay time of the scintillator light are not
considered.

The setup used for the calculation was an array with 162
elements with an inner radius of 10 cm. The thickness of the
crystals was chosen according to the attenuation length of the
scintillator. An aluminum beam pipe of 3 mm thickness was also
included. We assumed that the sample is 1 mm thick gold,
located at the center of the array.  This sample was irradiated
with neutrons in the energy range from 0.1 eV up to 20 MeV. The
energy spectrum of the neutrons was of the form 1/$E_n$ (see
section 6.1) and the flight path was 20 m long. TOF spectra for
events from neutron capture in the sample and for events from
neutrons which were scattered on the sample and captured in the
scintillator material were recorded separately. Various TOF cuts
corresponding to neutron energy decades between 0.1 eV and 1 MeV were
applied and the ratio between events caused by scattered and
captured neutrons was calculated. The neutron scattering-to-capture
ratio for these different bins varies from about 1:1 to more than
50:1. The response of the following 
scintillator materials was investigated by choosing the detector 
thickness such that the efficiency for capture events in the gold 
sample were equal in all cases.
\begin{itemize}
\item{ A BaF$_2$-array with a crystal thickness of 15 cm using the
natural composition of barium.}
\item{The same BaF$_2$-array but with the (hypothetical) assumption
that only $^{138}$Ba was contained in the scintillator. }
\item{A corresponding array of bismuth germanate (BGO) with a reduced 
thickness of 10 cm according to the higher $\gamma$-absorption. Since there are
no photon data for neutron capture on germanium in the GCALOR data base,
2 $\gamma$-rays with 25 $\%$ and 75 $\%$ of the full
energy were assumed to be emitted in each capture event}.
\item{CeF$_3$ represents an alternative solution requiring an 
intermediate crystal thickness of 13 cm}.
\item{Finally a C$_6$F$_6$ liquid scintillator tank with 120 cm diameter
was considered for comparison. However, the resolution in $\gamma$-ray energy
was so low that this possibility was not pursued any further.}
\end{itemize}
In all these simulations, cladding of the scintillators and mechanical support
structures were neglected. 

The resulting neutron sensitivities are compared in Table\,\ref{tab_sens}.
Since the C$_6$F$_6$ solution is excluded because the limited light collection
from large volumes deteriorates the resolution in $\gamma$-ray energy, one finds 
that BaF$_2$ and BGO are about comparable, but that CeF$_3$ is almost 10 times 
better at low neutron energies. However, this is the energy range where the 
scattering corrections are relatively small (see Fig.\,\ref{fig_ensim}). In the 
important region above 10 keV this advantage is no longer very significant, 
in particular if one considers the effect of good neutron absorbers 
(Figs.\,\ref{fig_absorb} and \ref{fig_abres}) and the possibility of background 
separation using energy cuts (50\% of the scattering events in BaF$_2$ appear at 
energies above 7.5 MeV). Hence, the better energy resolution makes BaF$_2$ 
still the scintillator of choice for a 4$\pi$ calorimeter.

\subsection{Sample radioactivity}

Background caused by the decay of a radioactive sample may present
a particular challenge. Usually, a decay deposits not more than about
1 MeV energy in the crystals whereas capture events deposit between
5 and 7 MeV. Therefore the problem is not deciding between single
background and capture events but rather is related to the high
radioactive count rate of a 1 mg sample with a lifetime between a
few days and several years, resulting in massive pile up effects
and dead time losses.

A listing of many isotopes of astrophysical interest is given in
Table\,\ref{tab_radio}. For a sample mass of 1 mg the average number of counts
in the detector system for an integration time of 150 ns (more than
enough for the fast component) and 2 $\mu$sec (slow component) is given in
columns 2 and 3, where we assume no shielding against radiation from the sample.
In this case the probability of an accidental coincidence with true
capture events would be unity for almost all of these isotopes. A lead shielding
of 5 mm thickness would be sufficient to solve this
background problem
(columns 4 and 5) except for a few cases. The corresponding
probabilities were calculated according to Poisson for the standard setup of 160 
crystals assuming a spherical lead shielding of 5 mm thickness so that 
$\gamma$-rays from the sample in the center have to pass at least 5 mm lead. 
For a few isotopes a 5 mm thick lead shielding is not
yet sufficient. Some of these, e.g. $^{85}$Kr, $^{170}$Tm, and $^{210}$Bi,
could still be measured if the lead shield is replaced by a 20 mm thick gold 
sphere \cite{HRK99}. Although it might not be practical because 
of its large (n,$\gamma$) cross section, gold was chosen for illustration 
due to its high atomic number and, compared with lead, its high density. 

\begin{table}
 \caption{ Simulated background from some radioactive isotopes of astrophysical 
interest (sample mass 1 mg, see text). Isotopes marked by an asterisk are decay products of the isotope listed 
before with lifetimes shorter than the lifetime of the parent nucleus. Due 
to decay equilibrium the decay rates of daughter and parent isotope are 
equal.
} 
   \label{tab_radio}
   \begin{tabular}{cccccc}
 Backing & \multicolumn{2}{c}{No shielding} & \multicolumn{2}{c}{5 mm Pb} & Radiation\\
         & 150 ns & 2 $\mu$s & 150 ns & P$_{coinc}$ & Energies in (keV) unless otherwise specified\\
    \hline
$^{79}$Se      & 0.38          & 5.2           & 0             & 0              & $\beta^-$: 150 \\
$^{85}$Kr      & 8.7           & 115           & 4.6           & 1              & $\gamma$: 514; $\beta^-$: 700\\
$^{90}$Sr      & 770           & 1 10$^4$      & 0             & 0              & $\beta^-$: 550 \\
$^{90}$Y$^*$   & 770           & 1 10$^4$      & 6.2           & 1              & $\gamma$: 1760; $\beta^-$: 2.3 MeV \\
$^{94}$Nb      & 1.1           & 14            & 1.0           & 0.6            & $\gamma$: 871+703; $\beta^-$: 470 \\
$^{106}$Ru     & 1.8 10$^4$    & 2.4 10$^5$    & 0             & 0              & $\beta^-$: 40 \\
$^{106}$Rh$^*$ & 1.8 10$^4$    & 2.4 10$^5$    & 3 10$^3$      & 1.0            & $\gamma$: 512; $\beta^-$: 3.6 MeV \\
$^{135}$Cs     & 7 10$^{-3}$   & 0.1           &  0            & 0              & $\beta^-$: 200 \\
$^{147}$Pm     & 13            & 180           & 0             & 0              & $\gamma$: 121; $\beta^-$: 220 \\
$^{151}$Sm     & 141           & 2 10$^3$      & 0             & 0              & $\gamma$: 20; $\beta^-$: 80 \\
$^{155}$Eu     & 1400          & 1.9 10$^4$    & 1.4 10$^{-3}$ & 1.4 10$^{-3}$  & $\gamma$: 87+105; $\beta^-$: 250 \\
$^{153}$Gd     & 2 10$^4$      & 2.6 10$^6$    & 1.5 10$^4$    & 1              & $\gamma$: 97+103; $\beta^+$: 500 \\
$^{163}$Ho     & 2.6           & 35            & 2.0           & 0.87           & $\beta^+$: 3 \\
$^{169}$Er     & 4.5 10$^5$    & 6.0 10$^6$    & 0             & 0              & $\gamma$: 110; $\beta^-$: 350 \\
$^{170}$Tm     & 6.9 10$^3$    & 8.0 10$^4$    & 0             & 0              & $\gamma$: 84; $\beta^-$: 1000 \\
$^{171}$Tm     & 3.3 10$^4$    & 4.4 10$^5$    & 0.7           & 0.5            & $\gamma$: 67; $\beta^-$: 100 \\
$^{175}$Yb     & 9.4 10$^4$    & 1.2 10$^6$    & 2.4 10$^4$    & 1              & $\gamma$: 396; $\beta^-$: 450 \\
$^{182}$Hf     & 1.2 10$^{-3}$ & 1.6 10$^{-2}$ & 0             & 0              & $\gamma$: 270; $\beta^-$: 160 \\
$^{182}$Ta$^*$ & 1.2 10$^{-3}$ & 1.6 10$^{-2}$ & 1.2 10$^{-3}$ & 1.2 10$^{-3}$  & $\gamma$: 1.1+1.2 MeV; $\beta^-$: 1.7 MeV \\
$^{185}$W      & 5.2 10$^4$    & 7.0 10$^5$    & 1.0 10$^{-5}$ & 1.0 10$^{-5}$  & $\gamma$: 125; $\beta^-$: 430 \\
$^{193}$Pt     & 0             & 0             & 0             & 0              & no $\gamma$ \\
$^{204}$Tl     & 66            & 880           & 0             & 0              & $\beta^-$: 800 \\
$^{210m}$Bi    & 3.1 10$^{-3}$ & 4.2 10$^{-2}$ & 0             & 0              & $\gamma$: 266; $\alpha$: 4.9 MeV \\
\end{tabular}
\end{table}

\begin{figure}
\includegraphics[width=.9\textwidth]{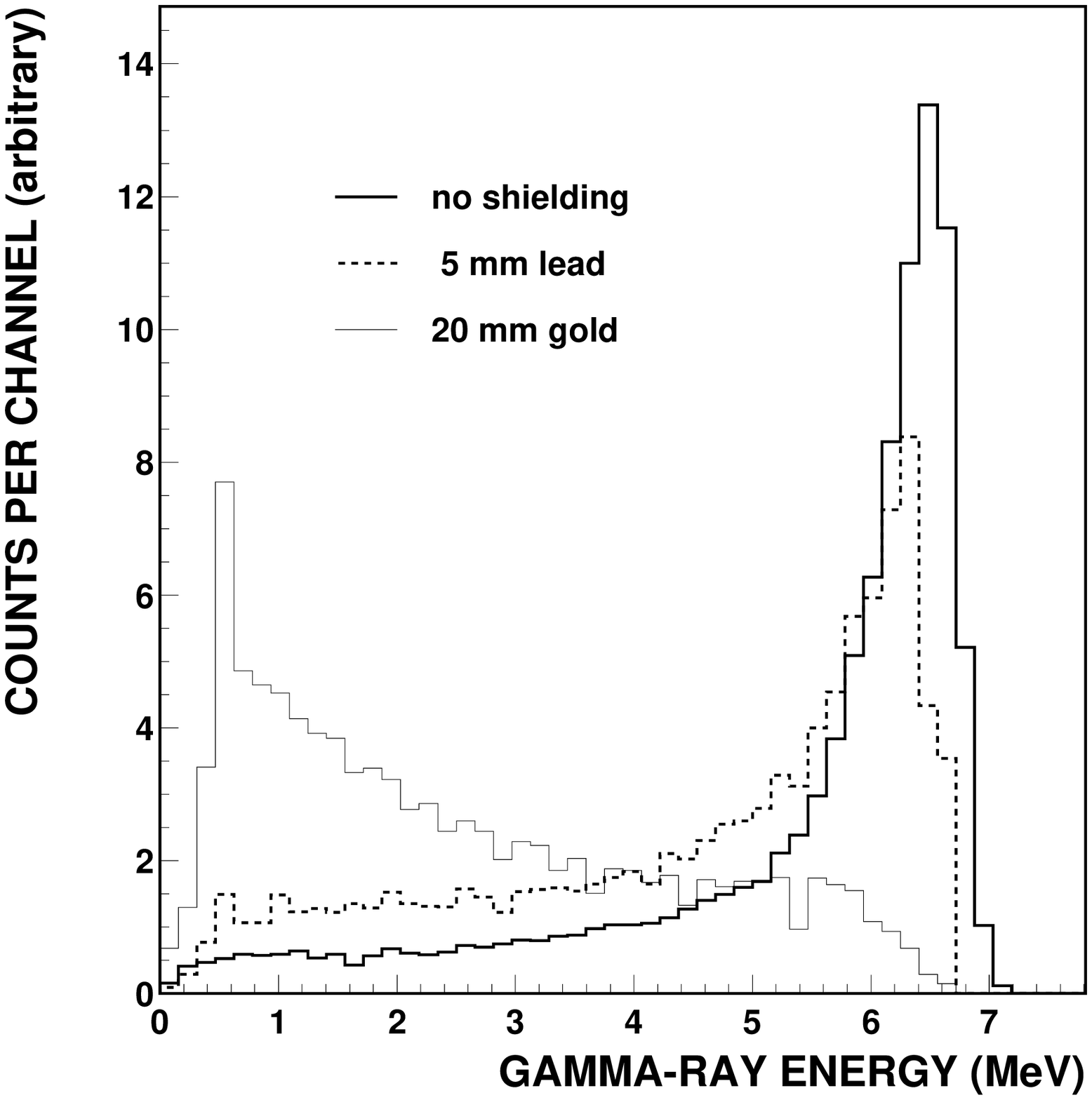}
\caption{Influence of various $\gamma$-shieldings on the $^{197}$Au$(n,\gamma)$ 
spectrum.
\label{fig_degr}}
\end{figure}

Shields of 5 mm lead and especially of 20 mm gold influence
significantly the $\gamma$-ray resolution of the detector as
shown for the detector response to $^{197}$Au$(n,\gamma)$ cascades,
in Fig.\,\ref{fig_degr}.

In spite of this degradation in energy resolution, the total efficiency 
for capture events is much less affected. In the ideal case considered, it 
drops from 100 \% (without  shielding) to 98.5 \% for a 5 mm lead absorber.
However, the massive gold shielding would result in a reduction to 54 \% 
and would imply that the full energy peak disappears completely.
But even in this extreme example the spectrum exhibits still enough 
events at high sum energies to allow for background discrimination.

\subsection{Neutron beam profile}

Collimation of the neutron beam for high energy neutrons (up to several
hundred MeV) is much more difficult than collimation of lower
energy neutrons.  To investigate the problem of a beam halo produced by
collimating the beam upstream of the sample, we assumed a beam core diameter of
30 mm (umbra), a penumbra of 80 mm diameter and an umbra/penumbra ratio of 2 x
10$^5$.  The umbra was assumed to cover a gold sample 30 mm in diameter with a
thickness of 100 $\mu$g/cm$^2$ for a total mass of 0.7
mg.  The penumbra, after passing the sample position, was incident on the
BaF$_2$.  Even with this very small beam halo, the halo background was
comparable to the neutron capture signal on the gold.  Good collimation of the
spallation neutron beam is therefore of great importance.

\section{Summary}
\noindent

We have considered a 4$\pi$ array of BaF$_2$ scintillators as a detector for 
neutron cross section measurements at a spallation neutron source.  Many design 
options have been included in the Monte Carlo GEANT simulations.

We find that this type of detector, which detects the total $\gamma$-ray energy 
following the capture event, has significant advantages over other detectors in 
that capture on the sample can be differentiated from other events simply by the 
Q-value of the reaction.  The energy resolution of the array must be good to 
take advantage of this essential feature.  Factors that would worsen the 
resolution, such as incomplete energy collection or using only a fraction of the 
light (such as just the fast component from BaF$_2$ crystals) or interaction of the 
$\gamma$-rays with 
components (such as the beam pipe, neutron absorbers, and aluminum and teflon 
reflectors around the scintillators) are quantified.  If further discrimination 
against background is required, the hit patterns on highly segmented detectors 
can be used.

Effects of neutrons scattered from the sample can be significant, especially for 
those samples that have large scattering-to-capture ratios.   The effects are 
particularly pronounced above 100 keV, where for most nuclei, the capture cross 
sections become quite small.  Although BaF$_2$ has a relatively small capture 
cross section, other scintillators such as CeF$_3$ and C$_6$F$_6$ can have even 
smaller cross 
sections, and their appropriateness for a given experiment should be kept in 
mind.  Segmentation of the BaF$_2$ was found to be a useful approach to 
identifying capture events from the sample and for discriminating against 
capture of scattered neutrons in the scintillator.

This study has not extended to data acquisition, where advanced techniques such 
as digital signal processing might be helpful especially when the instantaneous 
rate is high.

The detector designed here has many attractive features not only for neutron 
capture experiments but also for the study of $\gamma$-ray production in neutron-
induced fission and neutron inelastic scattering.

Acknowledgments - This work has benefited from the use of the Los Alamos Neutron 
Science Center at the Los Alamos National Laboratory.  This facility is funded by 
the U.S. Department of Energy and operated by the University of California under 
Contract No. W-7405-ENG-36. Three of us (M.H., F.K., R.R.) are greateful for the 
support of Los Alamos National Laboratory during several visits.

\newpage

\newpage

\newcommand{\noopsort}[1]{} \newcommand{\printfirst}[2]{#1}
  \newcommand{\singleletter}[1]{#1} \newcommand{\swithchargs}[2]{#2#1}

\end{document}